# Thermo-magnetic hysteretic properties resembling superconductivity in the normal state of $La_{1.85}Sr_{0.15}CuO_4$

M. Majoros[1] and C. Panagopoulos[1,2,*]

[1]*IRC in Superconductivity, University of Cambridge, Cambridge CB3 0HE, UK*
[2]*Cavendish Laboratory, University of Cambridge, Cambridge CB3 0HE, UK*

T. Nishizaki

*Institute of Materials Research, Tohoku University, Sendai 980-8577, Japan*

H. Iwasaki

*School of Materials Science, JAIST, Tatsunokuchi 923-12, Japan*

**Abstract**

We have performed detailed magnetic and thermal hysteresis experiments in the normal-state magnetization of $La_{1.85}Sr_{0.15}CuO_4$ single crystal. Using a combination of in-field and in-zero-magnetic-field measurements at different stages of thermal history of the sample, we identified subtle effects associated with the presence of magnetic signatures which resemble those below the superconducting transition temperature ($T_c$=36 K) but survive up to 250 K.

# I. Introduction

The magnetization of type II anisotropic superconductors displays thermal hysteresis below the irreversibility temperature due to the pinning of superconducting vortices [1]. Contrary to the conventional wisdom, recent systematic measurements on $La_{2-x}Sr_xCuO_4$ single and poly-crystalline samples revealed the presence of hysteresis in the temperature dependence of the low field magnetization up to a doping dependent characteristic temperature $T_s$, reaching a maximum of 290 K for $x$=0.10 [2]. The temperature, magnetic field, and crystallographic dependences of the onset and strength of the hysteresis were found to resemble fundamental properties of the mixed state. The magnetization results and correspondence with magneto-thermal transport experiments [3,4] opened the exciting possibility of either the presence of superconducting signatures, or some form of magnetism encouraging superconductivity above the superconducting transition temperature $T_c$. However, the nature of the currents responsible for the thermal hysteresis remains to be identified. To this aim we developed a novel experimental method which can be applied to study subtle magnetic effects in superconducting and magnetic materials in general. As we describe below this method has also the capability of distinguishing the intrinsic thermomagnetic hysteresis from possible contribution arising from extrinsic magnetic impurities. We have studied $La_{1.85}Sr_{0.15}CuO_4$ single crystals and the sample discussed here had a mass of 1.1 mg with onset $T_c$=36 K (transition width of 1.5 K). The samples were prepared and characterized as in ref. [5]. Our results indicate that the onset of the thermal hysteresis at $T_s$ is due to a transition from a fluctuating (at $T>T_s$) to a pinned ordered state ($T \leq T_s$), whose magnetic moment may be associated with superconductivity related long-lived persistent currents.

The paper is organized as follows. In Part II we briefly describe the experimental technique, in Part III we present the results on the thermal hysteresis effects observed by setting the applied magnetic field below and above $T_c$ (Part III.1 and III.2, respectively). In Part III.3 we show that $T_s$ is a transition temperature and in Part III.4 we report experiments showing the presence of magnetic flux conservation when crossing $T_c$ (both with decreasing and increasing temperature) in zero applied magnetic field. Part IV summarizes the present work.

**II. Experimental**

Measurements were carried out over a course of 12 months using a Quantum Design (MPMS-XL) SQUID magnetometer at 3 cm scan length after we first applied a zero magnetic field procedure to suppress the remnant field of its superconducting magnet down to ≤ ±1.5 G. No extra material was added to hold the sample and a sample-holder with no discontinuity was used so there was no background due to the sample holder – at least to the resolution of the XL-SQUID magnetometer. (The sample was held mechanically by two plastic straws placed inside an almost equal length outer straw supplied by Quantum Design.) It is the discontinuity which is responsible for the SQUID signal and therefore, in the experimental configuration with no discontinuity in the sample holder the measured signal reflects purely the sample's magnetization. As in our earlier work [2] the results reported here were confirmed several times over the 12 months course of the present experiments and with different scan lengths (4 and 6 cm), and heating, cooling times (0.5, 1, 2 and 4 K/min.). Note that the results shown here have been obtained by stabilizing the sample at every temperature the data was collected. We have been particularly careful to eliminate the possibility of thermal lags and external noise such as that coming from the mains. The raw data and SQUID response function were monitored for each data point.

As shown in ref. [2] the thermal hysteresis is stronger when the applied magnetic field $H||ab$ than when $H||c$. It is also for $H||c$ that the magnetic field dependence of the thermal hysteresis is weaker. Also the onset temperature of the hysteresis, $T_s$, decreases with increasing field, is smaller for $H||c$, and becomes negligibly small at $H>1$ kG. For these reasons in the present paper we performed all the experiments with $H||ab=100$ G.

**III. Results and discussion**

*1) Hysteresis effects by setting the applied magnetic field on at $T<T_c$*

The experimental procedure employed is as follows. First, we cooled the sample from 300 K to 6 K in zero-field. Once the sample was at 6 K, the applied magnetic field was set to 100 G and the magnetic moment, *m*, was measured first with increasing temperature up to 300 K (field warming – FW) (Fig. 1 – solid circles) and then with decreasing temperature back to 6 K (field cooling – FC) (Fig. 1 – solid squares). While at 6 K the magnetic field was set to zero and the sample was warmed up to 300 K (zero field warming – ZFW) (Fig. 1 – solid diamonds). The sample was then cooled back to 6 K in zero field (zero field cooling – ZFC) (Fig. 1 -main panel – open triangles).

Cooling the sample from above $T_c$ but in the presence of 100 G is known to result in homogeneously distributed trapped vortices at $T<T_c$ (Fig. 1 - inset – solid squares) [1]. By setting the applied field to zero at 6 K we observed the effect of induced currents which keep the field, vortices, trapped inside the sample and give rise to a paramagnetic moment below $T_c$ (Fig. 1 - inset – solid diamonds). As expected, these currents decrease with temperature. However, instead of vanishing at $T_c$ the moment survives at $T>T_c$ (Fig. 1 - main panel – solid diamonds). The presence of the positive moment, which was established at $T<T_c$ in order to keep the vortices trapped, up to $T_s$ may be taken as suggestive for the presence of trapped vortices above $T_c$ [2]. Next we cooled the sample from 300 K to 6 K in zero-field, and the data followed the "zero-line" (Fig. 1 - main panel – open triangles), indicating absence of any magnetization in the sample, as expected in an equilibrium paramagnetic state. We note that the existence of a straight "zero-line" as well as other extensive tests we perform on our samples [6], indicate that the above hysteresis effect cannot be due to traces of possibly undetected magnetic impurities.

The presence of trapped field at $T<T_s$ implies the presence of a thermal hysteresis. Indeed, this leads to the second observation of our experiments, *i.e.*, the thermal hysteresis at $T>T_c$. The sensitivity of the experiment allows us to see that the FW curve saturates at $T>T_s \approx 250$ K, at which point it also starts to be significantly noisier (Fig. 1 - main panel – solid circles). Notably, the FC curve, from 300 K to $T_s$ (Fig. 1 – main panel – solid squares), remained noisy too but basically reversible and with decreasing temperature at $T<T_s$, $m$ increased slightly and was always above the equilibrium background ($\approx 10^{-6}$ emu). This indicates that $T_s$ is not just a crossover temperature. Note that along the FW curve (solid circles) at $T_c<T<T_s$ the moments are lower than the equilibrium background at $T_s<T\leq 300$ K, indicating persistent diamagnetic currents may have survived as the sample crossed $T_c$.

*2) Hysteresis effects by setting the applied magnetic field on at $T>T_c$*

To investigate whether setting the applied field on at $T<T_c$ is essential for observing a hysteresis above $T_c$, we performed the following experiments. By warming the sample to 300 K in zero field, applying a magnetic field of 100 G at 300 K, and then cooling the sample again but only to 75 K, the data (Fig. 1 - main panel – open squares) traced the curve obtained previously for the original FC run (Fig. 1 - main panel – solid squares). Stopping at 75 K, decreasing the field to zero (as indicated by the big arrow pointing down in the main panel of Fig. 1) and subsequently warming the sample in zero field to 300 K, the data (Fig. 1 - main panel – open diamonds) traced the curve obtained for the ZFW run (Fig. 1 – main panel - solid diamonds). Returning from 300 K in zero field along the zero line (Fig. 1 - main panel – open triangles) and increasing the field to 100 G at 75 K (along the big arrow pointing up in the main panel of Fig. 1), the data (Fig. 1 - main panel – solid triangles) traced the FW curve obtained previously by increasing the field while the sample was at $T<T_c$ (Fig. 1 - solid circles). Moreover, increasing the applied field to 100 G, or decreasing it to zero and cooling or warming the sample in-field or in zero-field in the region 260 K$<T\leq$300 K resulted in a zero line (for 0 G) (Fig. 1 - main panel – crosses) and in noisy data for 100 G - on a paramagnetic background of $\approx 10^{-6}$ emu (Fig. 1 – main panel, crossed squares). Furthermore, the data obtained at 100 G in this temperature region (Fig. 1 - main panel - open squares) were always noisier than the data obtained in the temperature region $T<T_s$. These tests show that the same results are obtained at $T>T_c$ as long as the field is increased or decreased below $T_s$.

The main panel in Fig. 1 also indicates that the in-field and in-zero-field data have different backgrounds on which the hysteresis is superimposed: one for $H=0$ (Fig. 1 - triangles) and one for $H=100$ G (Fig. 1 - crossed squares). In order to compare the levels of the hystereses we corrected the in-field data on their paramagnetic background at $T_s<T\leq 300$ K ($10^{-6}$ emu) and re-plotted them in Fig. 2. The result we obtain is a nearly symmetric picture at $T_c<T<T_s$ (Fig. 2 - solid circles and diamonds). Interestingly, this picture resembles the behavior observed at $T<T_c$ (Fig. 1- inset), indicating a common origin in the moments giving rise to the thermal hysteresis below and above $T_c$. Furthermore, the in-field data are noisier than the zero-field data, and the noise is higher at $T_s<T\leq 300$ K. Decreasing the field to zero from the FC data at 75 K we observe an increase in $m$ along the short arrow pointing up in Fig. 2 (open squares → open diamonds), suggesting the presence of persistent paramagnetic currents induced by the change in the field (similarly as in Fig. 1, inset, solid diamonds).

*3) Tests showing that $T_s$ is a transition temperature*

The presence of a thermal hysteresis when $H$ is applied at $T<T_s$, and the increased noise above $T_s$ indicate the latter is not a mere crossover. To examine whether an actual transition occurs at $T_s$, we studied the behavior of the hysteresis when we return from $T<T_s$. Figure 3 depicts properties of such partial magnetic and thermal hystereses loops: We have warmed the sample to 150 K after applying a magnetic field of 100 G at 6 K. The sample was then FC down to 75 K (Fig. 3 – upper solid triangles). By decreasing the applied field to zero at 75 K, $m$ dropped along the big arrow - starting from the upper solid triangles and pointing down to lower solid triangles in Fig. 3. The sample was then ZFW to 300 K, with $m$ ending on the curve which is below that obtained by the ZFW process - starting either at $T<T_c$ ($T=6$ K) (solid diamonds) or at 75 K (Fig. 1 – main panel - open diamonds). Moreover, Fig. 3 shows that by lowering $H$ to zero at 75 K on the FW curve (solid circles), $m$ dropped along the big arrow pointing from solid circles down to solid right-angle triangles, and it is slightly above the zero line. It has also a weak but distinct temperature dependence. Therefore, it is only when we decrease the applied field to zero while the sample is at $T>T_s$ that the true zero moment is observed. Hence $T_s$ represents a transition from a fluctuating to a pinned ordered state in which long-lived weak persistent currents may exist - depending of course on the magnetic history of the sample. Let us note that the drop of the magnetic moment at 75 K to nearly zero as the field was decreased (Fig. 3 – big arrow pointing from solid circles down to solid right-angle triangles), differs from observations in conventional spin glasses, where the moments decay very slowly [7].

*4) "Flooding" experiments*

The question arising next, is how to collectively understand these and earlier [2] observations. Although there is no theoretical interpretation of our results, given the absence of bulk superconductivity above $T_c$, and based on the striking resemblance of the hystereses at $T<T_c$ and $T_c<T<T_s$, one suggestion may be the presence of superconducting "vortices" at $T>T_c$ [2]. In fact signatures for their possible presence in the normal state have been observed in magneto-thermal transport and local magnetic imaging experiments in LSCO [3,4,8]. As discussed previously [2] such presence would explain many of our observations, in particular the similarities in the hystereses below and above $T_c$, and the increased moment in Fig. 2 (short arrow pointing up). Overall, the proposed picture here would be similar to the critical state model below $T_c$. Alternatively, the data may be governed by a mechanism incorporating magnetic domains, for example in the form of droplets or rivers [9].

Irrespective of the nature of the domain structure (vortices or not) causing the thermal hysteresis, the fundamental question which needs to be experimentally addressed is whether this form of magnetism co-operates or competes with superconductivity. Already, our earlier work has shown that $T_s(x) \sim T_c(x)$, suggesting a cooperative relation ($x$ is the carrier concentration level) [2]. To address this question in some more detail here we "flooded by superfluid" this "magnetism" by first FW the sample from 6 K to 300 K in 100 G, and then FC to 6 K. While at 6 K, we decreased the applied magnetic field to zero, and ZFW the sample to 75 K (Fig. 4 – solid diamonds). From that point, but still in zero field, we cooled the sample to 6 K (Fig. 4 - crosses) and warmed it again to 300 K (Fig. 4 - crossed squares). The data obtained by cooling and warming the sample in the range of 75 K – 6 K – 75 K was (a) reversible, (b) the magnetic moment did not depend on temperature, and (c) the magnetic moment crossed smoothly from the normal to the superconducting state and vise versa. A similar behavior was observed also when the sample was ZFW to 150 K and then cooled to 6 K and warmed to 300 K (Fig. 4 – open squares and solid triangles).

This reversibility shows clearly a return-point-memory effect, indicating the crossing through $T_c$ is a smooth evolution from one state ($T<T_c$) to another ($T>T_c$), and there a common origin for the magnetic moment responsible for the hysteresis above and below $T_c$. This indicates a magnetic flux conservation effect (Faraday effect) rather than a Meissner effect. To see it more clearly, we show the raw data (Fig. 4 - upper inset) of the sample ZFW to 75 K, then cooled to 6 K, and finally warmed to 300 K. The magnetic moment corresponding to the zero line (open triangles) is negative at $T>T_c$, and there is a jump to positive values at $T_c$ as the temperature drops below $T_c$.

This effect is similar to the decrease of the applied magnetic field to zero when the sample is in the superconducting state (e.g., at $T=6$ K) after it was FC from 300 K in 100 G (Fig. 1, inset, solid rectangles – solid diamonds). The non-zero magnetic moment of the zero line at $T>T_c$ is a consequence of a trapped magnetic field in the superconducting magnet of the SQUID magnetometer used to perform the measurements. Because the magnetic moment $m=[\mu_o^{-1}V/(1-N)](B-\mu_oH_a)$ (in SI units) ($V$ – sample volume, $0<N<1$ – demagnetizing factor, $B$ – magnetic flux density, $H_a$ – applied magnetic field) is negative, it means that $B<\mu_oH_a$. A jump in the magnetic moment to positive values at $T<T_c$ can in principle be caused by the Meissner effect, if the trapped magnetic field of the superconducting magnet $H_a$ is negative. When the field $B=0$, $m=[\mu_o^{-1}V/(1-N)](-\mu_oH_a)>0$. Now, because the trapped field $H_a$ of the magnet does not change, cooling the sample from 75 K, when $B$ in the sample is higher than that on the zero line, should give at least the same positive magnetic moment as for the zero line, or lower (because of a possible partial screening), but not higher as observed in the present experiment (crosses and crossed squares in the upper inset of Fig. 4).

It is difficult to identify the exact effects of a trapped field on the "zero-line" because we measure only $m$, which is a difference of two unknown quantities – $B$ and $H_a$. Furthermore, the value of the trapped field is random. More importantly however, the data corrected on the background due to the trapped field (by its subtraction) (Fig. 4 - the main panel) show no jump when crossing $T_c$ - in both directions. This means that the flux in the sample is conserved, and no Meissner effect is observed. The magnetic flux conservation points towards the presence of a Faraday-effect. This means that when crossing $T_c$, at $T<T_c$ superconducting screening currents are induced, which keep the field inside the sample unchanged. Because the trapped field $H_a$ of the superconducting magnet does not change with the temperature of the sample (these are two independent systems), the only possibility is that the Faraday effect is triggered by a change in the magnetic flux density $B$ inside the sample at $T=T_c$. In other words, the paramagnetism of the sample at $T>T_c$ must start to diminish (its increase is less probable because the magnetic moment does not change with temperature as we start to cool down the sample from e.g. 75 K or warm it up – Fig. 4, crosses and crossed squares) giving rise to the induced surface superconducting currents, which then keep the paramagnetic moment in the bulk unchanged. By warming the sample above $T_c$ these macroscopic induced superconducting currents disappear at $T_c$ because no bulk superconductivity is observed in the normal state of the sample. This suggests a cooperative behavior of the magnetism and superconductivity - in the sense that the paramagnetism of the sample at $T>T_c$ tends to disappear in the superconducting state as the temperature drops below $T_c$ i.e., the paramagnetic moments tend to change their orientation. This normal state paramagnetism may be caused either by pinned vortices, stripes, or by some other form of magnetic domains. However, there must exist some interaction between the supercurrent and these domains at $T_c$, leading to a change of their magnetic moments. A similar memory effect is observed also at $T<T_c$ (Fig. 4 – the lower inset). These results suggest that the magnetic structures above and below $T_c$ do not compete and the currents responsible for the measured moment above $T_c$ (Fig. 4 – solid diamonds) are those responsible for the moment below $T_c$ (Fig. 1 – inset – solid diamonds).

**IV. Summary**

We report the magnetization of a $La_{1.85}Sr_{0.15}CuO_4$ single crystal under various history protocols. Using a special combination of in-field and in-zero-magnetic-field cooling and warming experiments we showed that the observed hysteresis effects cannot be caused by spurious magnetic impurities. We have demonstrated that the same thermal hysteresis effects in the normal state are obtained not only when the applied magnetic field is increased or decreased at $T<T_c$, but also when it is increased or decreased at $T>T_c$ - for as long as the temperature is below $T_s$. We showed that $T_s \approx 250\ K$ represents a transition from a fluctuating to a pinned ordered state ($T \leq T_s$), in which long-lived weak persistent currents may exist. In zero applied magnetic field, crossing $T_c$ - both by warming and cooling - results in a smooth reversible transition indicating that the paramagnetic moments at $T>T_c$ interact with the supercurrent at $T<T_c$. The unprecedented similarity found in the thermal and magnetic history behavior of the measured magnetization below and above $T_c$ [Fig. 1 (inset) and Fig. 2, Fig. 4 (upper inset), respectively] adds credence to our earlier suggestion [2] for a common cause of the hysteretic behavior in the two temperature regions.

M.M. acknowledges the AFRL/PRPS Wright-Patterson Air Force Base, Ohio, for financial support. C.P. and the experimental work in Cambridge were supported by The Royal Society.

\* Corresponding author, e-mail: *cp200@hermes.cam.ac.uk*

[1] J.R. Waldram, *Superconductivity of metals and cuprates.* Institute of Physics (1996).

[2] C. Panagopoulos, M. Majoros, and A. P. Petrovic, Phys. Rev. B **69,** 144508 (2004).

[3] Z. A. Xu, N. P. Ong, Y. Wang, T. Kakeshita, and S. Uchida, Nature (London) **406,** 486 (2000).

[4] Y. Wang, Z. A. Xu, T. Kakeshita, S. Uchida, S. Ono, Y. Ando, and N. P. Ong, Phys. Rev. B **64,** 224519 (2001).

[5] H. Iwasaki, F. Matsuoka, and K. Takigawa, Phys. Rev. B **59**, 14624 (1999).

[6] X-ray diffraction, micro-Raman, scanning probe microscopy, and electron probe microanalysis indicated no traces of impurities in our samples, at least to the level of 90ppm. Our earlier studies on different single crystals and polycrystals, with both same as well as different Sr concentrations (prepared in three different laboratories and with two different preparation methods) showed similar hysteresis effects [2]. Furthermore, as we discussed in Ref. 2, for any given sample we confirm our results on smaller pieces cut from the original measured piece. Some of the smaller pieces were also polished and re-measured to confirm the absence of possible surface effects. Separate experiments like those discussed in Fig. 1 were performed on pure Fe, and Fe-contaminated samples. The results showed that the data obtained by cooling the sample in zero magnetic-field do not follow the straight "zero-line" but the magnetic moment increases with decreasing temperature. Moreover, the ratio of the width of the hysteresis at 100 K to the magnetic moment at 300 K is one order of magnitude lower for the iron and iron-contaminated samples than that obtained from Fig. 1. Furthermore, in our samples (both here and in Ref. 2) the onset of the thermal hysteresis has strong doping dependence, crystallographic anisotropy, and the values of $T_s$ are incomparable to those expected for any composition of iron, iron oxide, or other ferromagnetic impurities in general. These experiments exclude the possibility that the hysteresis effects reported here and earlier [2] are due to possible undetected magnetic impurities. Of course there is always the possibility that a tiny amount of impurity phase might still be present and be beyond detection by any of the aforementioned chemical, spectroscopic and thermodynamic tests we preformed. If that is the case, then that impurity will have

to follow the stringent systematic trends discussed in ref. 2 and here. To the best of our knowledge there is no such magnetic phase.

[7] K. H. Fischer, J. A. Hertz, *Spin glasses*. Cambridge University Press (1991).

[8] I. Iguchi, T. Yamaguchi, and A. Sugimoto, Nature (London) **412,** 420 (2001).

[9] S. A. Kivelson, E. Fradkin, and V. J. Emery, Nature (London) **393,** 550 (1998).

**Figure captions**

**FIG. 1** Magnetic moment *m* versus temperature *T* for a $La_{1.85}Sr_{0.15}CuO_4$ single crystal in applied magnetic field $H||ab$=100 G. (●) – field applied at 6 K and sample warmed in field up to 300 K. (■) – sample cooled in field from 300 K to 6 K. The main panel and inset depict the high temperature ($T>T_c$) and low temperature regions, respectively. (◆) – field decreased to zero at 6 K and sample warmed in zero field up to 300 K. (△) – sample cooled in zero field from 300 K to 6 K. (□) – field applied at 300 K and sample cooled in field to 75 K. (◇) - field decreased to zero at 75 K (along the big arrow pointing downward) and sample warmed in zero field up to 300 K. (▷) – sample cooled in zero field from 300 K to 6 K. (⊠) – field applied at 300 K and sample cooled in field to 260 K. (**x**) – field decreased to zero at 260 K and sample warmed in zero field (ZFW) up to 300 K. (◢) – sample cooled in zero field from 300 K to 75 K. (▲) – field applied at 75 K (along the big arrow pointing upward) and sample warmed in field up to 300 K. (▼) – sample cooled in field from 300 K to 6K.

**FIG. 2** Magnetic moment *m* versus temperature *T* for a $La_{1.85}Sr_{0.15}CuO_4$ single crystal in applied magnetic field $H||ab$=100 G. Data from field warming and field cooling runs are corrected on their equilibrium background (a reversible part between 250 K and 300 K, value $10^{-6}$ emu – see Fig. 1). The meanings of the symbols are the same as in Fig. 1.

**FIG. 3** Magnetic moment $m$ versus temperature $T$ for a $La_{1.85}Sr_{0.15}CuO_4$ single crystal in applied magnetic field $H||ab$=100 G. (●) (■) (◆) (△) (□) (▷) (◇) – their meanings are the same as in Fig. 1. (▲) – field applied at 6 K, sample warmed in field up to 150 K (these points are not shown in the figure) and then cooled in field to 75 K. (▼) – field decreased to zero at 75 K (along the big arrow starting from upper solid triangles and pointing down to lower solid triangles) and sample warmed in zero field to 300 K. (◣) – field applied at 6 K, sample warmed in field up to 75 K (along the solid circles) then field set to zero (along the big arrow pointing from solid circles down to solid right-angle triangles) and sample zero-field warmed up to 300 K.

**FIG. 4** Magnetic moment $m$ versus temperature $T$ for a $La_{1.85}Sr_{0.15}CuO_4$ single crystal in applied magnetic field $H||ab$=100 G. (◆) (△) – their meanings are the same as in Fig. 1. To arrive to the crosses (x) the following procedure was performed: A 100 G field was applied at 6 K, then the sample was warmed in field up to 300 K and consequently cooled in field to 6 K. The field was then decreased to zero at 6 K and the sample warmed in zero-field up to 75 K. Then the sample was cooled down to 6 K in zero-field. The crosses shown in the figure capture the latter part of the procedure. The data shown as crossed squares (⊞) capture the part where the sample was warmed again from 6 K and in zero-field up to 300 K. To arrive to the open squares (□) (main panel) the following procedure was performed: A 100 G field was applied at 6 K, then the sample was warmed in field up to 300 K and consequently cooled in field to 6 K. The field was then decreased to zero at 6 K and the sample warmed in zero-field up to 150 K. Then the sample was cooled down to 6 K in zero-field. The open squares shown in the figure capture the latter part of the procedure. The data shown as solid triangles (▲) capture the part when the sample was warmed again from 6 K and in zero-field up to 300 K. In the upper inset all the symbols have the same meaning as in the main figure, but the data are not corrected for the background. In the lower inset (♦) – captures the part described in Fig. 1 for the same symbols but at $T<T_c$, whereas (□) and (♦) at 6 K $\leq T \leq$ 12 K represent a thermal cycle performed to test the memory of the moment below $T_c$.

Figure 1

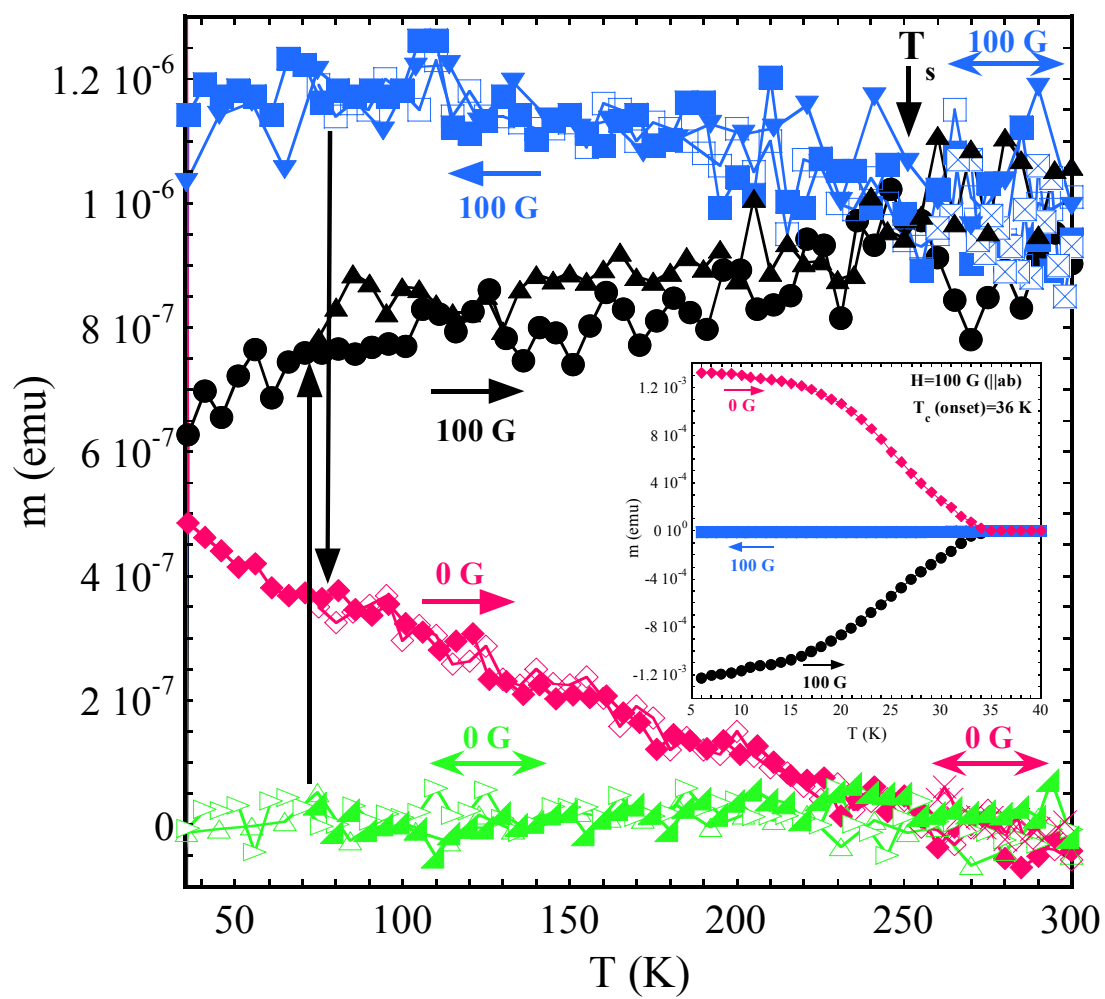

Figure 2

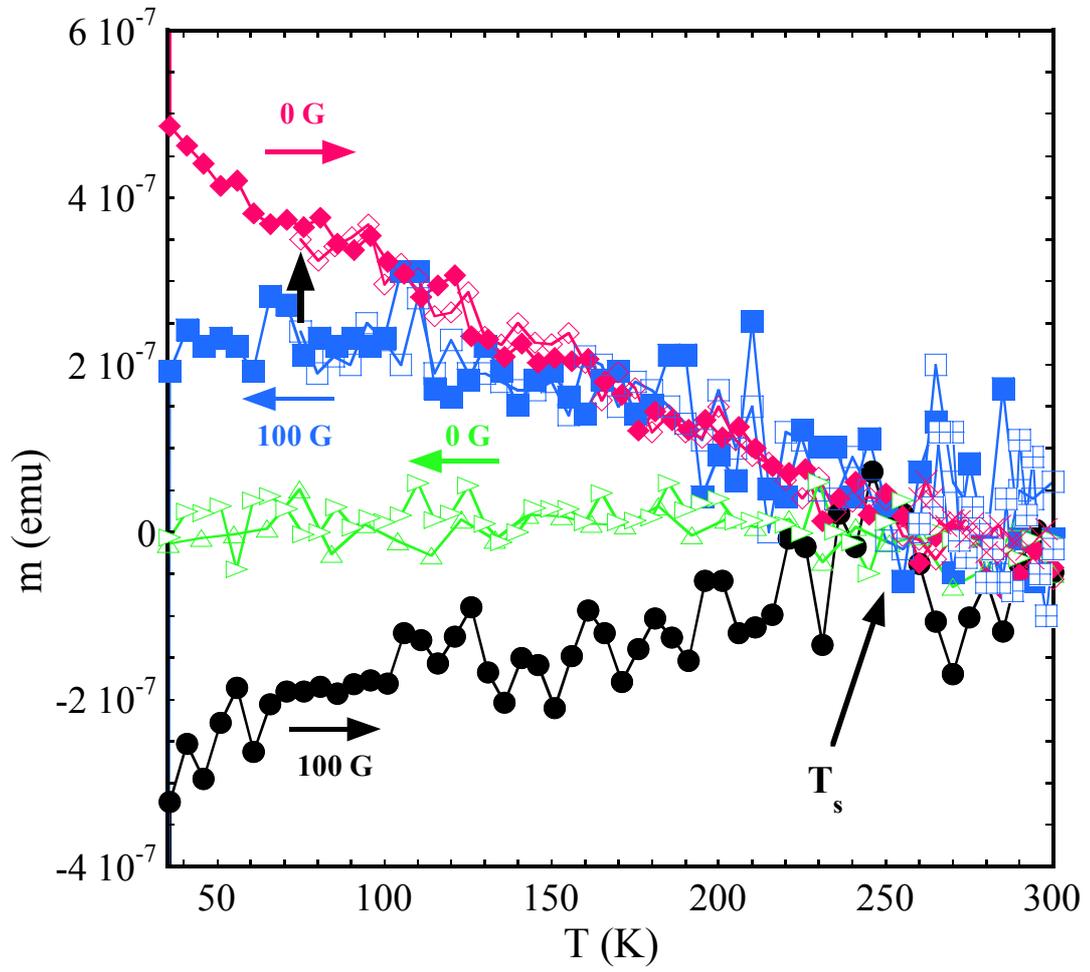

Figure 3

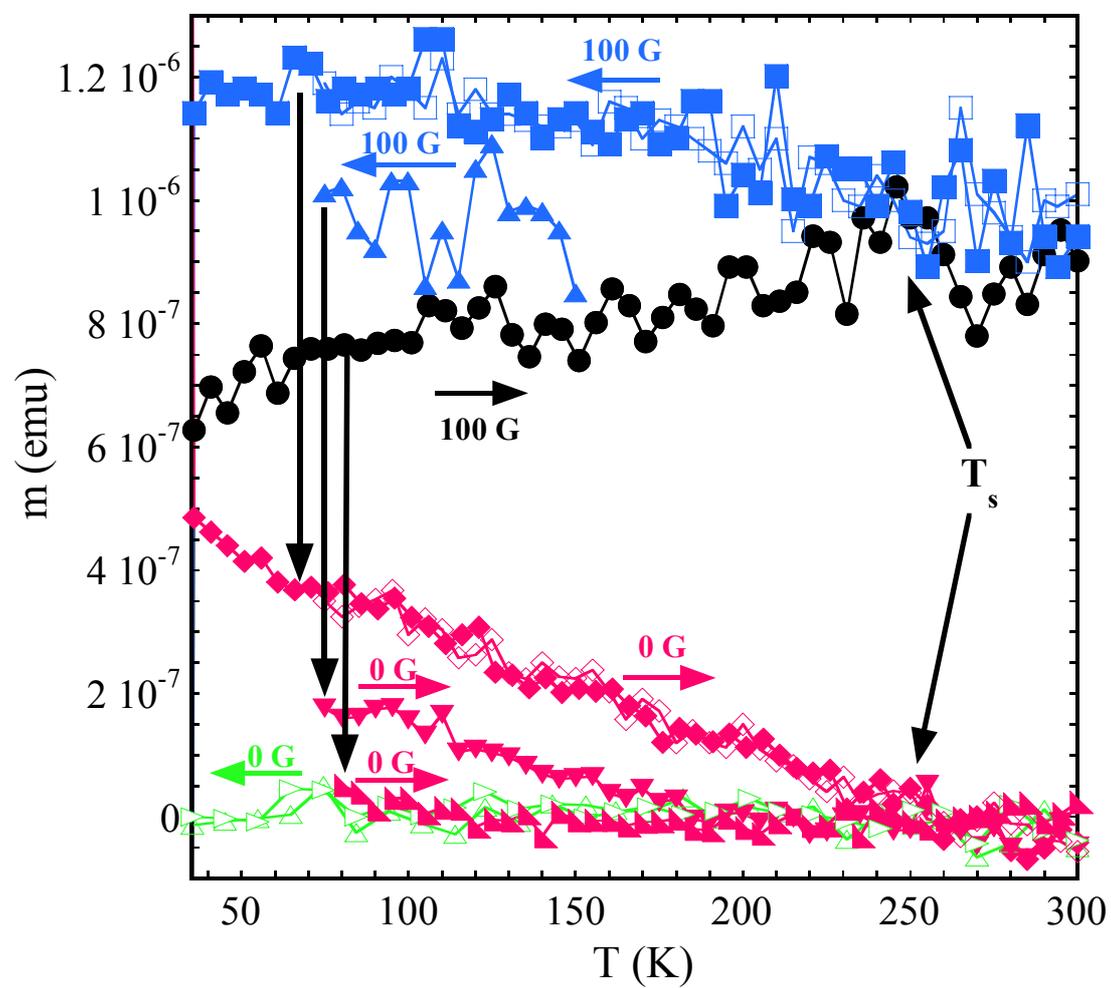

Figure 4

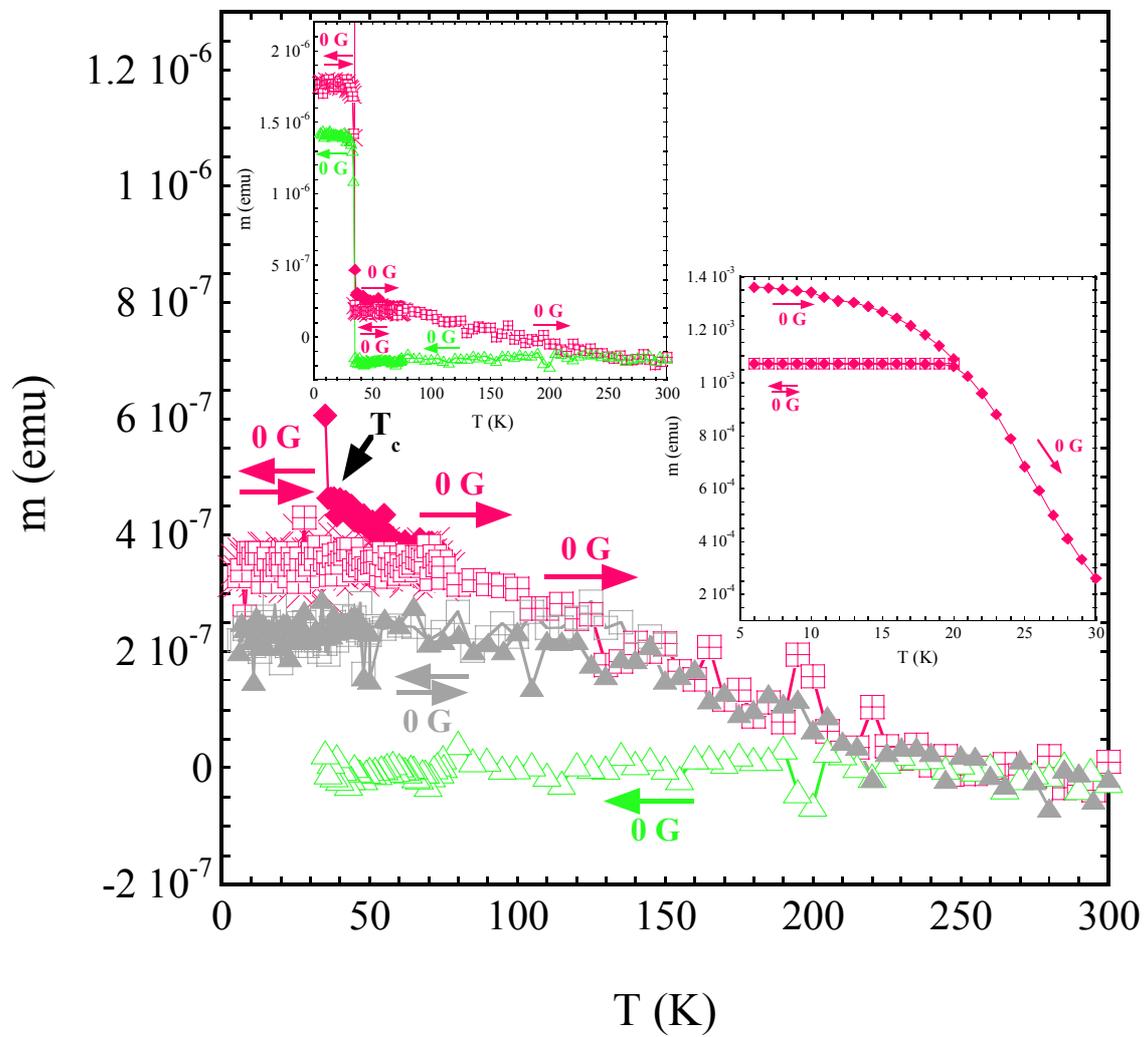